\documentclass[lettersize,journal]{IEEEtran}
\usepackage{amsmath,amsfonts}
\usepackage{algorithmic}
\usepackage{array}
\usepackage[caption=false,font=normalsize,labelfont=sf,textfont=sf]{subfig}
\usepackage{textcomp}
\usepackage{stfloats}
\usepackage{url}
\usepackage{verbatim}
\usepackage{graphicx}
\usepackage{cite}

\usepackage{svg}
\newcommand{\orcid}[1]{\href{https://orcid.org/#1}{\includesvg[width=10pt]{orcid}}}

\usepackage[bookmarks=false]{hyperref} 
\begin{document}
\title{A Submillimeter-Wave FMCW Pulse-Doppler Radar to Characterize the Dynamics of Particle Clouds}

\author{Tomas~Bryllert,~\IEEEmembership{Member, IEEE},
Marlene~Bonmann, 
and~Jan~Stake,~\IEEEmembership{Senior Member,~IEEE} 

\thanks{Manuscript received January 1\textsuperscript{st}, 2023; revised February 23\textsuperscript{rd}, 2023. This work was supported in part by the Swedish Foundation for Strategic Research (SSF) under the contract ITM17-0265.}
\thanks{Tomas Bryllert, Marlene Bonmann, and Jan Stake are with the Terahertz and Millimetre Wave Laboratory, Chalmers University of Technology, SE-412 96 Gothenburg, Sweden. (e-mail: \mbox{tomas.bryllert@chalmers.se}; \mbox{marbonm@chalmers.se}; \mbox{jan.stake@chalmers.se})}}

\markboth{IEEE Transactions on Terahertz Science and Technology,~Vol.~13, No.~x, X~2023}%
{How to Use the IEEEtran \LaTeX \ Templates}

\maketitle

\begin{abstract}
This work presents a 340-GHz frequency-modulated continuous-wave (FMCW) pulse-Doppler radar. The radar system is based on a transceiver module with about one milli-Watt output power and more than 30-GHz bandwidth. The front-end optics consists of an off-axis parabola fed by a horn antenna from the transceiver unit, resulting in a collimated radar beam. The digital radar waveform generation allows for coherent and arbitrary FMCW pulse waveforms. The performance in terms of sensitivity and resolution (range/cross-range/velocity) is demonstrated, and the system's ability to detect and map single particles (0.1–10\,mm diameter), as well as clouds of particles, at a 5-m distance, is presented. A range resolution of $\sim$1\,cm and a cross-range resolution of a few centimeters (3-dB beam-width) allow for the characterization of the dynamics of particle clouds with a measurement voxel size of a few cubic centimeters. The monitoring of particle dynamics is of interest in several industrial applications, such as in the manufacturing of pharmaceuticals and the control/analysis of fluidized bed combustion reactors.
\end{abstract}

\begin{IEEEkeywords}
FMCW, pulse-Doppler, radar, remote sensing, sensors, submillimeter waves, terahertz systems, transceivers
\end{IEEEkeywords}

\section{Introduction}
\IEEEPARstart{F}{or} many industrial applications, such as in the manufacturing of pharmaceuticals \cite{BAWUAH2021}, or energy conversion using fluidized bed reactors \cite{KOORNNEEF_2007}, the industrial process involves particles or powders dispersed in a process reactor. It is necessary to monitor the particle dynamics to maintain the process quality and to gain insights regarding the process. Therefore, measuring the particle concentration and the local particle velocities at a high update rate and high spatial resolution is desirable. Ideally, these quantities should be measured ex vivo without inserting any physical probes into the reactors so that introducing measurement sensors does not alter the processes. In particular, this is required in harsh process environments \cite{Zankl2015}. Frequency-modulated continuous-wave (FMCW) range-Doppler radar operating at center frequencies ($f_c$) within the submillimeter wave range \cite{Siegel2002} of the electromagnetic spectrum offers a realistic opportunity to provide the desired information.

Compared to other contactless measurement methods using visible or infrared light \cite{WERTHER1999, FRAKE1997}, the submillimeter wavelength range allows more penetration depth into dense particle clouds \cite{Appleby2007} and is less sensitive to contaminations on the reactor access windows. The radar technique also allows for Doppler processing, which reveals information about the velocities of the particles \cite{BONMANN2023}. Compared with more traditional radar techniques in the microwave and millimeter wave region \cite{Kueppers2022}, there are a few properties that favor submillimeter waves \cite{Cooper2014}:
\begin{itemize}
\item{Short wavelengths ($\lambda$) result in higher sensitivity for detecting smaller particles since the radar cross-section of particles in the Rayleigh regime scales as $\lambda^{-4}$;}
\item{Wide bandwidth and, thereby, a higher range resolution. For example, a 30-GHz bandwidth results in a theoretical range resolution of 5\,mm;}
\item{The cross-range resolution for a fixed antenna size, typically limited by the access window size in an actual application, improves with high frequency since the diffraction-limited resolution scales with $\lambda$.}
\end{itemize}

 Several FMCW radars for high-resolution, 3D imaging have been presented with center frequencies above 300\,GHz \cite{Sheen2010, Cheng2018, Cooper_2008, Robertson2018}. FMCW radars using MMIC-transceivers based on SiGe technology have also been demonstrated in the millimeter wave region \cite{Thomas2019}, including promising performance up to 480\,GHz \cite{Mangiavillano2022}. Still, submillimeter-wave transceivers, with a high dynamic range at room temperature, require diode technology \cite{Mehdi2017, stake2017}. \textcolor{black}{Submillimeter wave FMCW radars that use range-Doppler processing, i.e., that allow measuring not only the range but also the velocity of a target, are rare.} Cooper et al. \cite{Cooper2014} reported a FMCW range-Doppler radar system at 660\,GHz, demonstrating the range-Doppler concept's feasibility at submillimeter wave frequencies \textcolor{black}{but with few details on sensitivity to different targets or the origins of the noise floor in the range-Doppler images.}

\begin{figure}
\centering
\includegraphics[width=8.8cm,keepaspectratio]{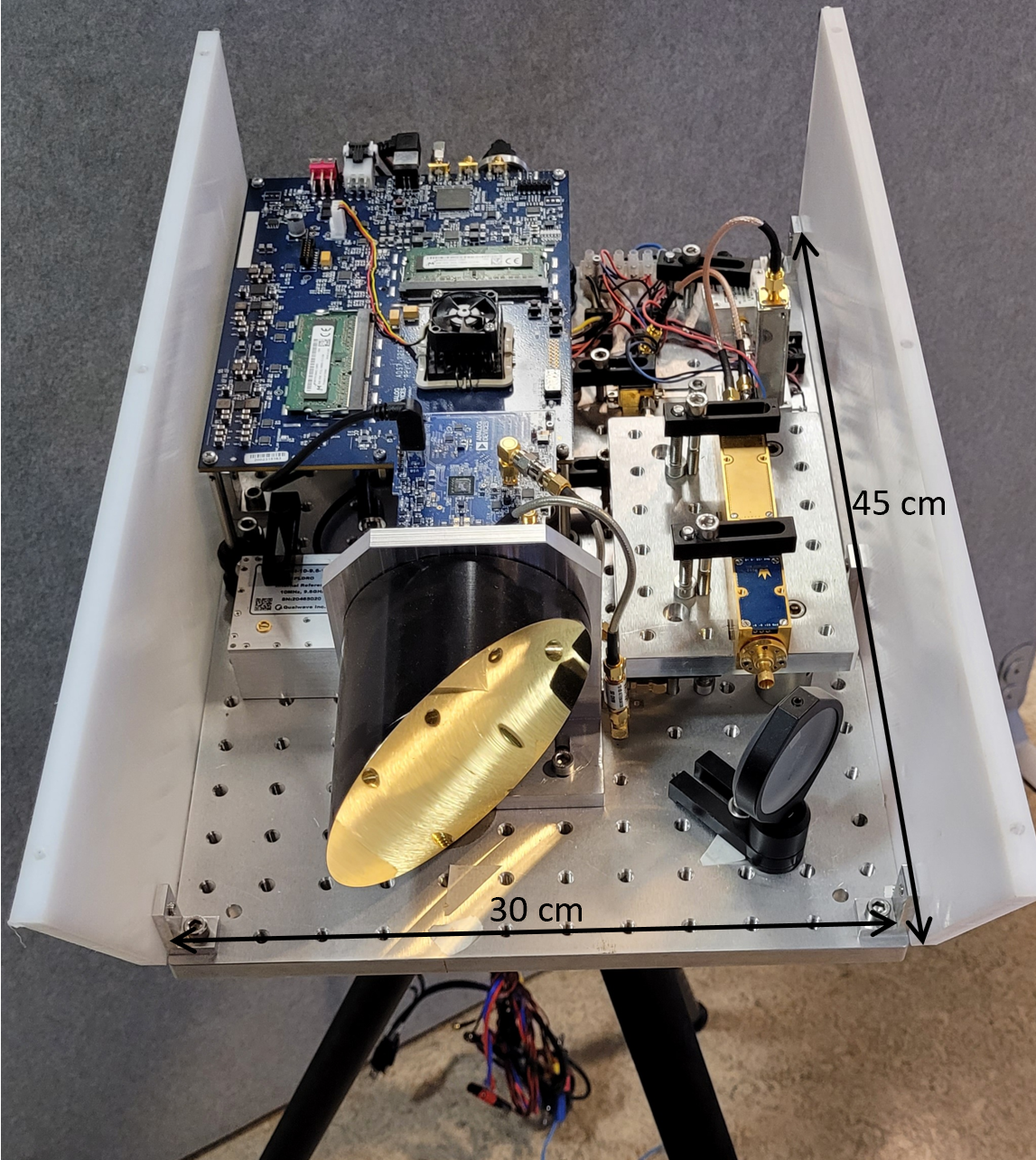}
\caption{Photograph of the radar system. The front-end optics and electronics are mounted on a base plate together with analog and digital baseband circuitry.}
\label{fig1}
\end{figure}

This work presents the implementation of a FMCW pulse-Doppler radar based on a 340-GHz transceiver module with 30-GHz bandwidth \cite{Dahlback2016}. A digital waveform generator controls the system. The transceiver module provides an acceptable trade-off between performance and hardware complexity, resulting in a relatively compact tripod-mounted radar design, as shown in Fig. \ref{fig1}. The form factor allows easy implementation in industrial scenarios. The performance of the transceiver modules and their application in a 3D imaging radar was presented in \cite{Robertson2018}. Here the implementation of the coherent pulse generation and signal processing to realize range-Doppler radar operation are explained, together with the resulting radar system's noise- and resolution performance. Furthermore, the ability of the radar to detect single particles with diameters ranging from 100\,$\mu$m – 500\,$\mu$m is demonstrated. \textcolor{black}{The capability of the velocity measurements is demonstrated by comparing the measured range-Doppler profile of a falling metal-sphere with known weight and diameter to the standard free-fall model}. The results demonstrate that the performance of the radar system is highly suitable for the suggested industrial scenarios.

\section{Method}
\subsection{Radar electronics and optics}
\begin{figure*}
\centering
\includegraphics[width=17cm,keepaspectratio]{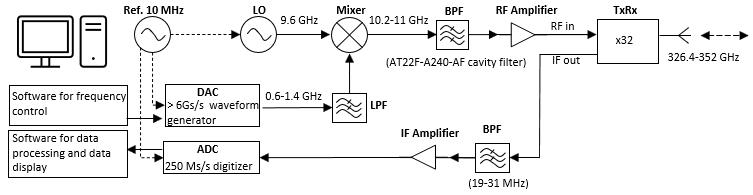}
\caption{Schematic block diagram of the 340-GHz FMCW pulse-Doppler radar.}
\label{fig2}
\end{figure*}

Fig. \ref{fig2} shows a schematic block diagram of the 340-GHz FMCW range-Doppler radar. The system architecture is a frequency up-converted, frequency multiplied FMCW radar.
A few hardware details deserve to be highlighted: The digital waveform generator is an FPGA-controlled arbitrary waveform card \textcolor{black}{(ADS7-V2EBZ from Analog Devices)} with 4\,Gb of useful memory and a maximum sampling rate of $>$6\,Gs/s of which 4\,Gs/s is used in the current work. The card can write $>$100\,ms of 1-GHz bandwidth waveform data directly from memory. This means that, in a coherent pulse-Doppler processing interval (CPI), typically much shorter than 100 ms, an arbitrary pulse train of FMCW pulses can be transmitted – and then repeated. Multiple FMCW waveforms can therefore be interleaved, addressing different parts of the system bandwidth (323 – 357\,GHz) within a coherent processing interval. This capability can 
 be used to extract frequency-resolved (spectroscopic) information from the scene in an efficient way. This feature is not used in the presented performance demonstrations. The baseband chirp, typically 1-GHz bandwidth, generated by the digital hardware, is centered at 1 GHz. This signal is up-converted to the X-band using frequency mixing and a 9.6-GHz local oscillator (LO) and is then passed on to the transceiver unit. 
The front-end 340-GHz  Schottky diode circuit is designed to operate as a frequency multiplier (x2) and sub-harmonic mixer - thereby simultaneously operating as a transmitter and receiver. The transceiver's LO chain consists of an InGaAs pHEMT active frequency multiplier MMIC (x8) developed by Gotmic AB and a 170-GHz Schottky diode frequency doubler. The GaAs Schottky barrier diode circuits were fabricated in the Nanofabrication Laboratory at Chalmers university of technology. Initially, the complete transceiver module was developed for a 16-channel, high frame-rate, imaging radar \cite{Robertson2018} by Wasa Millimeter Wave AB, and is described in detail in \cite{Dahlback2016}. \textcolor{black}{Thus, the transceiver unit multiplies the X-band chirp by a factor of 32 for a total final bandwidth of 32\,GHz and transmits the signal, now centered at $\sim$340\,GHz. The radar echoes are received back in the transceiver and are mixed on the outgoing signal straight down to the baseband using a balanced configuration \cite{Bryllert2013}.} 

At the output of the transceiver unit, a circular horn from Custom Microwave Inc is used as a feed antenna for the optical system. This feedhorn illuminates a 4" off-axis parabolic mirror from Edmund Optics with an effective focal length of 6". The optical system results in a collimated radar beam.

The digital hardware on the receiver side consists of an eight-channel (1 channel is used), 250-Ms/s digitizer from National Instruments \textcolor{black}{with 14-bit resolution}. The digitizer is controlled by an FPGA, which gives deterministic timing control. The digitizer card (PXIe format) integrates with a PC controller via a PXIe bus allowing for real-time signal processing and display. The waveform card, the analog-to-digital converter (ADC), and the local oscillator run from a common 10-MHz reference resulting in a fully coherent system.

\subsection{Radar signal processing}
\begin{figure*}
\centering
\includegraphics[width=17cm,keepaspectratio]{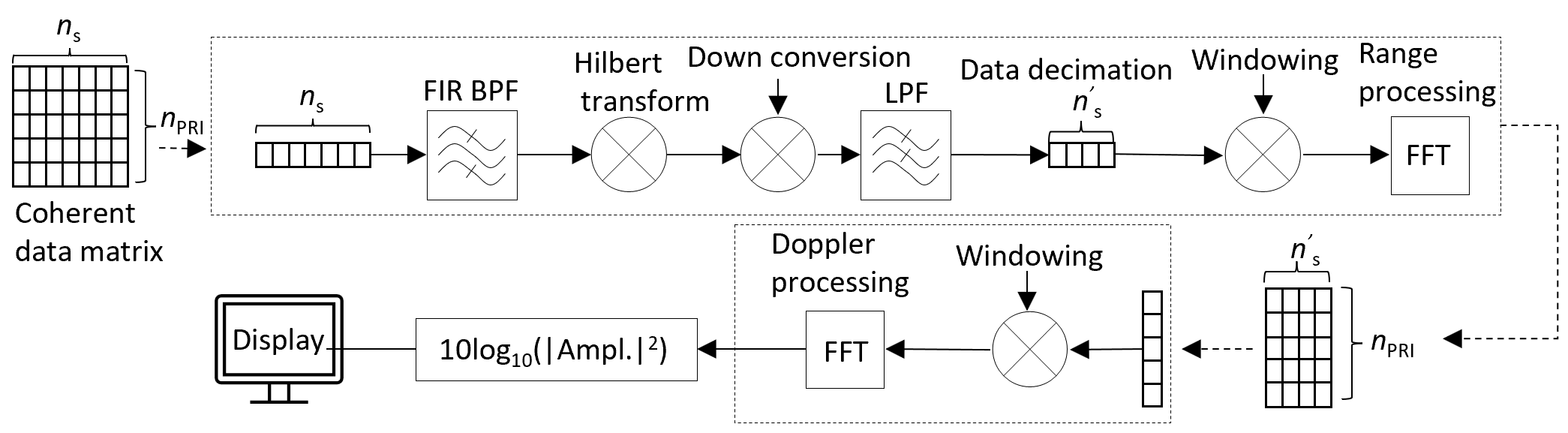}
\caption{Schematic block diagram of the digital signal processing steps.}
\label{fig3}
\end{figure*}

Typical radar parameters used in the experiments presented in this work are:
\begin{itemize}
\item{\textcolor{black}{Center frequency, \textcolor{black}{$f_c$} = 340\,GHz;}}
\item{Pulse bandwidth, \textcolor{black}{$BW$} = 32\,GHz;}
\item{Pulse time, \textcolor{black}{$t_p$} = 41\,$\mu$s;}
\item{Pulse repetition interval, \textcolor{black}{$PRI$} = 102.4\,$\mu$s or 51.2\,$\mu$s;}
\item{Number of pulses coherently processed, \textcolor{black}{$n_{PRI}$} = 128;}
\item{Target distance, \textcolor{black}{$R$ = 4 to 6\,m.}}
\end{itemize}
\textcolor{black}{These parameters set the theoretical limits to the maximum unambiguous range ($R_{max}$) and maximum unambiguous velocity ($v_{max}$), as well as the range-and velocity resolutions. With $c$ being the speed of light in air, $R_{max} = PRI\times c/2\approx 15\,$km and $v_{max} = c/(2\times PRI\times f_c) = 4.3$\,m/s. The range resolution is $\Delta R = c/(2\times BW)=5\,$mm. This is also the size of the range bins when no zero-padding is used in the range FFT of the radar signal processing, explained further below. 
In the Doppler dimension, the frequency resolution ($\Delta f_D$) is set by the coherent integration time ($t_c=PRI \times n_ {PRI}$) to $\Delta f_D=1/t_c$; this translates to a velocity resolution via the Doppler relation: $f_D= 2v/\lambda$, where $v$ is the radial velocity of the target. The velocity resolution is then: $\Delta v = \Delta f_D \times \lambda / 2 = 1 / t_c \times \lambda/2$ = 3.8 cm/s in the present work.}
Fig. \ref{fig3} shows a block diagram of the signal processing. The data matrix format that is coherently processed is of the form: (nr of samples per pulse, $n_s$) $\times$ ($n_{PRI}$). 
After down-conversion in the transceiver, the received baseband (IF) signal is \textcolor{black}{filtered with a 19 – 31\,MHz analog bandpass filter matching the target distance of interest. The signal is then digitized, digitally filtered with a 20 - 30\,MHz finite impulse response bandpass filter (FIR BPF)}, converted to IQ format with the help of the Hilbert transform, down-converted to complex baseband, and decimated by a factor of 16 to ~1.5 $\times$ Nyquist limited sampling (15.625\,Ms/s IQ), 
with: $n_s'$  = 640.
In reality, several samples at the beginning and the end of each waveform are discarded (due to low-frequency ringing), leaving 590 samples instead of 640. This also reduces the used bandwidth from 32\,GHz to 29.5\,GHz.
Both the pulse compression in range and the Doppler processing can be done using Fourier transforms in FMCW pulse-Doppler radar, which means that the signal processing can be done with a 2D fast Fourier transform (FFT) over the coherent data matrix – with appropriate \textcolor{black}{Hanning} windowing functions  and digital filters. The output displayed for the radar user is the logarithm of the squared amplitude of the radar signal in a range-Doppler map. 

\subsection{Radar characterization and evaluation}
To demonstrate the performance of the radar system in terms of the noise floor, range and velocity resolution, and small particle detection, the following measurements were conducted: noise floor measurements, range and Doppler resolution, detection of small particles, and velocity measurements of a free-falling metal sphere. 
To study the origin of the noise floor in zero-Doppler and at finite Doppler frequency, the noise floor was measured without a target under four different conditions: First, with ADC only; second, with ADC together with IF amplifiers; third, ADC with IF amplifiers and a 10.1\,GHz continuous wave (CW) signal driving the transceiver, fourth, ADC with IF amplifier and a chirp signal driving the transceivers.
Additionally, the noise floor as a function of target strength was measured. Different radar cross sections (RCSs) were achieved by placing a corner reflection at different positions in the radar beam. 

 Increasing the number of pulses per CPI, with other radar parameters fixed, the S/N for a target should increase linearly with the number of pulses (integration time) if the target and the radar system remain coherent and if the noise is uncorrelated with the radar signal. To verify this, a radar measurement on a static, corner reflector target was performed with $n_{PRI}$= 16, 32, 64, 128, 256, and 512 per CPI.
 
Three metal beads with a 2-mm diameter were glued onto a string and positioned at a 5\,m distance to demonstrate the radar system's range resolution. The target with three beads on a string was positioned so that all beads were illuminated by the radar beam and angled so that the beads were separated in range by approximately 3\,cm. 
Another radar measurement was performed to display the velocity resolution while gently tapping the string to make it vibrate.

To investigate the radar system's ability to detect small particles, the radar beam is folded with a flat metallic mirror to be directed vertically upwards. A transparent plastic box was placed directly above the folding mirror to collect the particles. This way, different test materials could be dropped straight into the radar beam. This experiment used 2-mm and 10-mm diameter metal beads, 500-$\mu$m diameter quartz sand, and 100-$\mu$m spherical glass beads.

\textcolor{black}{The ability to measure velocity ($v$) is demonstrated by comparing the measured position and speed of a free-falling metal sphere of known diameter (10\,mm) and weight ($m=4$\,g) to a well-established, analytical free-fall model. Letting the metal bead drop towards the radar, it moves vertically under gravity and quadratic air resistance ($\propto v^2$)}. Solving Newton's second law of motion, the velocity ($v$) and position ($x$) with time ($t$) are then described by 
\begin{subequations}\label{freefall}
\begin{align}
v=v_t\tanh{(t/\tau)}\label{freefall:A}\\
x=x_0-v_t\tau\ln{(\cosh{(t/\tau)})}\label{freefall:B}
\end{align}
\end{subequations}

\noindent with the terminal velocity $v_t=\sqrt{(2mg/(A\rho_{air}c_d))}$ and the characteristic time $\tau=v_t/g$, where $g$ is the gravity of Earth, $m$ is the mass of the metal bead, $\rho_{air}$ is the air density at normal temperature pressure, $A$ is the metal beads cross-section, $c_d$ is the drag coefficient (here 0.47 for a sphere \cite{miller_bailey_1979}), and $x_0$ is the initial position.

\section{Results}

\subsection{Noise performance}
\begin{figure}
\centering
\includegraphics[width=8.8cm,keepaspectratio]{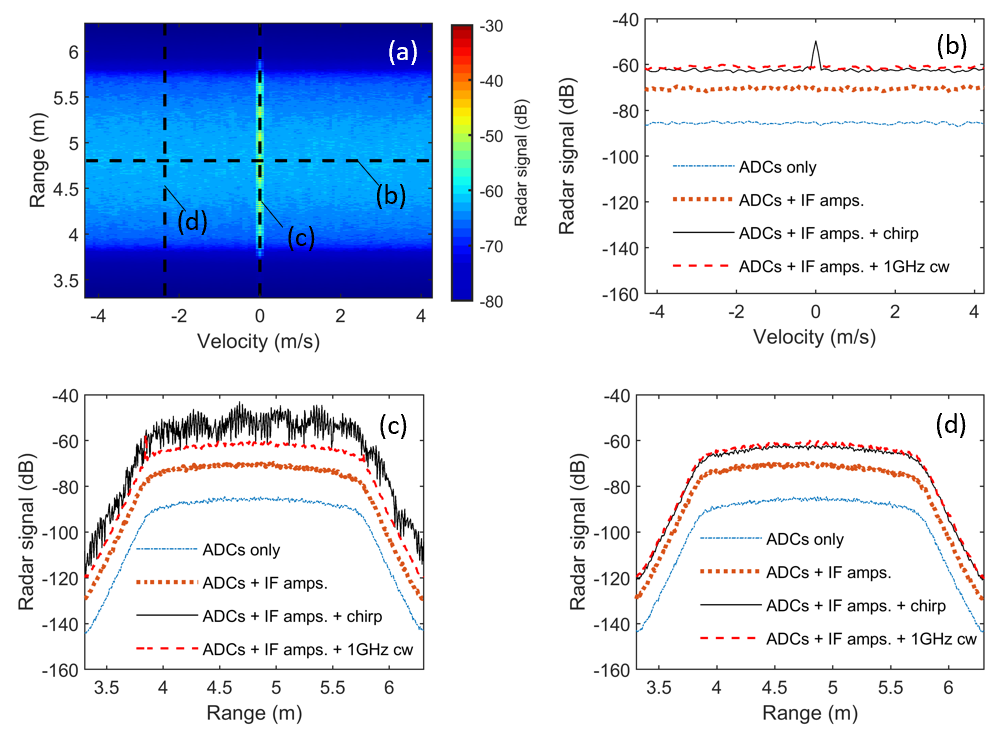}
\caption{Noise floor in the range-Doppler map. (a) General view of the noise floor with the cuts that are presented in (b-d) indicated. (b) Constant range cut. (c) Constant velocity cut at zero-Doppler. (d) Constant velocity cut at finite Doppler.}
\label{fig4}
\end{figure}

Fig. \ref{fig4} shows the noise floor at different hardware settings and different cuts through the range-Doppler map as indicated in Fig \ref{fig4}(a). No target is used in these measurements, aiming to demonstrate the origins of the noise floor for the radar.

Ideally, the noise floor in the whole range-Doppler map should be set by thermal noise, deteriorated by the loss and noise figure of the front-end electronics, and scaled by the IF amplification. The transceiver unit trades noise performance for simplicity though. Using the same balanced pair of Schottky diode circuits for the final stage frequency multiplication and subharmonic homodyne down-conversion to baseband \cite{Dahlback2016}, the transceiver unit can be made quite compact – at the cost of excess noise. 
The excess noise comes in two shapes – through a conversion loss in the subharmonic mixing that is worse than would be the case in a dedicated mixer and through excess amplitude modulated noise from the LO (FMCW chirp) that mix into the IF side of the transceiver (despite the balanced configuration). In addition, Fig. \ref{fig4}(c) shows that excess noise is generated in zero-Doppler from driving the RF hardware with short (40\,$\mu$s) chirps with high bandwidth\textcolor{black}{, which was also observed in \cite{Cooper2020}}. The cost of the excess noise is acceptable, though, since S/N is generally sufficient in the application scenarios that are evaluated. 

\begin{figure}
\centering
\includegraphics[width=8.8cm,keepaspectratio]{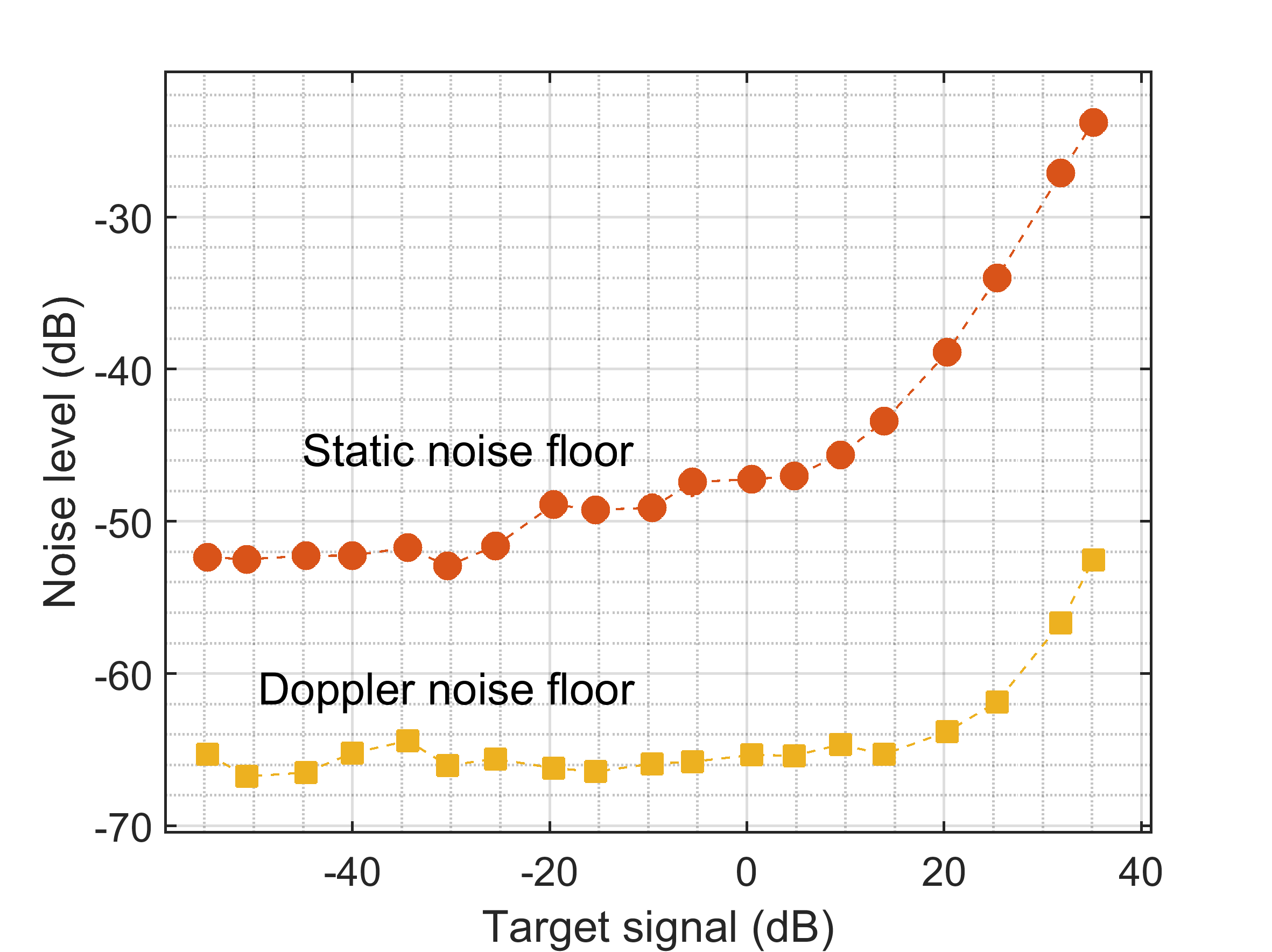}
\caption{The noise floor in zero-Doppler and at a finite Doppler frequency as a function of target strength (RCS).}
\label{fig5}
\end{figure}

Fig. \ref{fig5} shows how the noise floor for zero-Doppler and finite Doppler is affected by the strength (RCS) of a static target. The noise floor is calculated as the mean when averaging over relevant range bins within the IF filter bandwidth (excluding the range-bin with the target response). The noise floor in zero-Doppler is not random noise but the result of side lobes and amplitude/phase modulation of the waveform, as well as multiple reflections in the RF hardware. These effects are not seen at a finite Doppler frequency since the sidelobe/modulation/reflection pattern is identical from pulse to pulse and, therefore, only appears in the zero-Doppler bin. At strong target returns, the noise floor increases at finite Doppler frequencies but then as a general increase of the noise floor in the whole range-Doppler plane - indicating that this noise increase originates in the actual noise of the RF-carrier.

\begin{figure}
\centering
\includegraphics[width=8.8cm,keepaspectratio]{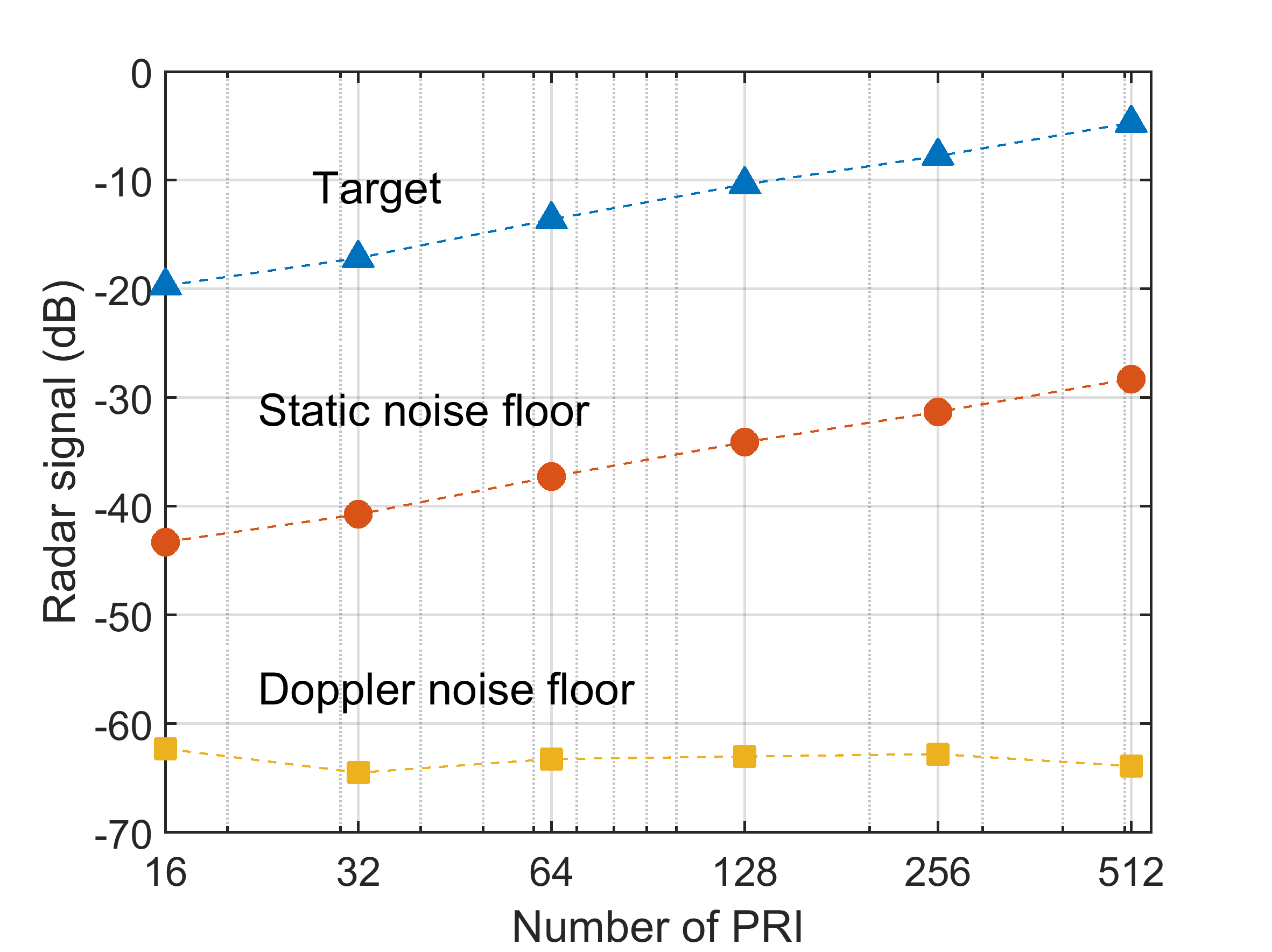}
\caption{S/N as a function of the number of pulses used in the coherent processing. The target was a static corner cube.}
\label{fig6}
\end{figure}

Fig. \ref{fig6} shows the radar signal of a static target and the noise floor versus the number of pulses per CPI.
 The S/N, when comparing the target signal with the Doppler noise floor, increased linearly as expected. Thus verifying that the target and the radar system remain coherent and the noise is uncorrelated with the radar signal. As discussed above, the noise in zero-Doppler (the static noise floor) originates from the radar signal, meaning no improvement in S/N in zero-Doppler is seen at longer integration times.
 
\subsection{Range resolution, small particle detection, and velocity measurement}
\begin{figure}
\centering
\includegraphics[width=8.8cm,keepaspectratio]{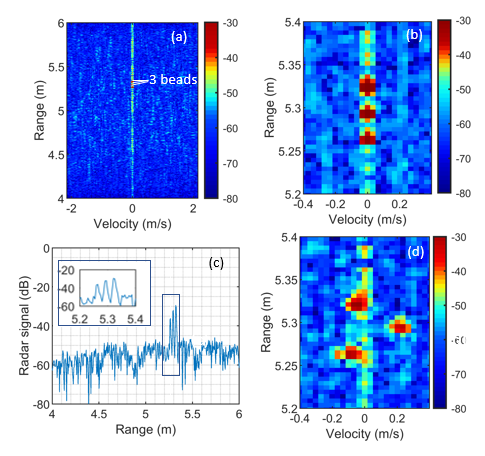}
\caption{Range and velocity resolution. (a-c) Shows a radar measurement of three beads with 2-mm diameter, demonstrating that the beads are resolved in range when positioned 3\,cm apart in the range direction. The S/N is approximately 20\,dB. (d) The string is vibrating, moving the beads in different directions and resulting in small Doppler shifts.}
\label{fig7}
\end{figure}

Fig. \ref{fig7} shows that the three metal beads with 2-mm diameter are clearly separated in the radar measurement with the signal peaks measured to be 3\,cm and 3.1\,cm apart and are visible with a S/N of approximately 20\,dB. \textcolor{black}{The reason for the beads to span several bins, despite being only 2\,mm in size is mainly because of
spectral widening from the Hanning windows that are used in the signal processing both in range and Doppler. The windows reduce spectral leakage and side lobes at the cost of widening the main lobe in Doppler and range.} When lightly tapping the string, the beads are also separated in Doppler due to the fine Doppler resolution of 0.04\,m/s per Doppler bin.

\begin{figure}
\centering
\includegraphics[width=8.8cm,keepaspectratio]{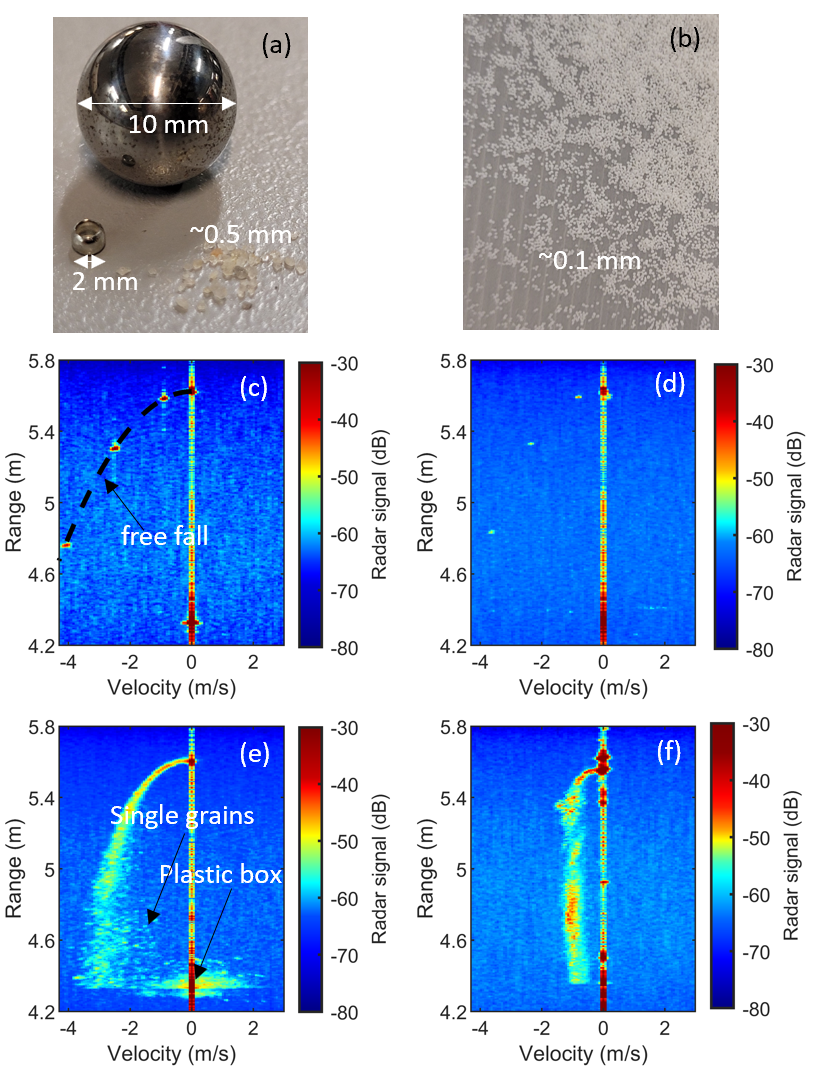}
\caption{Measurement of falling objects at a 5-m distance. (a-b) Show the photographs of the tested materials. The time-integrated range-Doppler image of (c) a 10-mm diameter falling metal bead, (d) a 2-mm diameter metal bead,  (e) a few pinches of 500 $\mu$m sand grains, and (f) a few pinches of 100 $\mu$m glass spheres. (c) Shows the predicted trajectory from the free-fall model (\ref{freefall}).}
\label{fig8}
\end{figure}

Figures \ref{fig8}(a-b) show photographs of the materials used for testing the radar system's ability to detect small particles. For each material, Figures \ref{fig8}(c-f) show the corresponding range-Doppler maps integrated over several CPI. This way, one can see the acceleration of the 2-mm diameter metal bead and the 10-mm diameter metal sphere toward the radar. Each detection corresponds to a separate CPI, or “frame”, of the radar with a frame rate of 6.2\,frames/s. For 500-$\mu$m diameter sand grains and 100-$\mu$m diameter glass spheres, the integrated particle stream over several pinches of particles is clearly visible. Fig. \ref{fig8}(e) shows clear detection of single sand grains. At a 4.3\,m distance, the sand grains hit the plastic box and bounced to a stop. The deflection from the plastic box appears as positive Doppler velocity. In conclusion, all tested materials could be detected with significant S/N at a 5-m distance, proving the radar instrument's suitability to monitor particle clouds' dynamics. 

Fig. \ref{fig8}(c) includes the predicted trajectory for the 10-mm diameter metal sphere from the free-fall model (\ref{freefall}), which indicates that the measurement agrees very well with the theory, thus supporting the velocity measurement of the radar. 
\textcolor{black}{Table \ref{comparison} compares the specifications of submillimeter-wave radars in terms of center frequency, bandwidth, output power, and technology, showing that the radar system presented here has comparable bandwidth, i.e., range resolution, to most of the other systems. However, only the radar system in \cite{Cooper2014} was operated in range-Doppler mode with a chirp bandwidth of 3\,GHz stating a range-and velocity resolution of 5\,cm and 0.2\,m/s, respectively. The results above demonstrate up to ten times better resolution of the radar system in this work.}

\begin{table*}
\begin{center}
\caption{Comparison of submillimeter-wave radars}
\label{comparison}
\begin{tabular}{| c | c | c | c | c | c |}
\hline
Center frequency & Bandwidth & Output power & Comment & Technology & Reference\\
(GHz) & (GHz) & (mW) &  &  &\\
\hline
350 & 19 & 4 & FMCW & Schottky diode & \cite{Sheen2010} \\
\hline
675 & 30 & 0.5 & FMCW pulse-Doppler & Schottky diode & \cite{Cooper2014} \\
\hline
340 & 29 & 0.6 & FMCW & Schottky diode & \cite{Reck2015} \\
\hline
332 & 16 & 0.2 & FMCW, MIMO & Schottky diode & \cite{Cheng2018} \\
\hline
340 & 30 & 1 & FMCW & Schottky diode & \cite{Robertson2018} \\
\hline
383 & 80 & 8 & FMCW & mHEMT & \cite{Baumann2022} \\
\hline
480 & 55 & 0.06 & FMCW & SiGe & \cite{Mangiavillano2022} \\
\hline
\textbf{340} & \textbf{30} & \textbf{1} &\textbf{FMCW pulse-Doppler} &\textbf{Schottky diode} & \textbf{This work} \\
\hline
\end{tabular}
\end{center}
\end{table*} 

\section{Conclusions}
We have presented a 340-GHz frequency-modulated continuous-wave pulse-Doppler radar. The performance of the radar is described and shown to follow what is expected from theoretical predictions. The instrument's sensitivity and resolution, both in the spatial domain and in Doppler velocity, are adequate to map the dynamics of particle clouds. This is demonstrated by performing radar measurements on free-falling particles with grain sizes down to  100-$\mu$m diameter. The mapping of particle clouds is relevant in many industrial applications, such as in the manufacturing of pharmaceuticals or energy conversion using fluidized bed reactors. \textcolor{black}{In addition, measuring cloud dynamics for meteorological applications \cite{Cooper2021} using terahertz, FMCW pulse-Doppler radar is an emerging field \cite{Cooper2020}.} Future work will demonstrate the radar technique in these applications.

\section*{Acknowledgment}
The authors would like to thank Mats Myremark for machining mechanical parts for the measurement setup; Vladimir Drakinskiy for his help with the fabrication of the front-end terahertz circuits; Divya Jayasankar for valuable feedback on the manuscript and help with \LaTeX. The devices were fabricated and measured in the Nanofabrication Laboratory and Kollberg Laboratory, respectively, at Chalmers University of Technology, Gothenburg, Sweden.

\bibliographystyle{IEEEtran}
\bibliography{IEEEfull,bibl}

\begin{IEEEbiography}[{\includegraphics[width=1in,height=1.25in,clip,keepaspectratio]{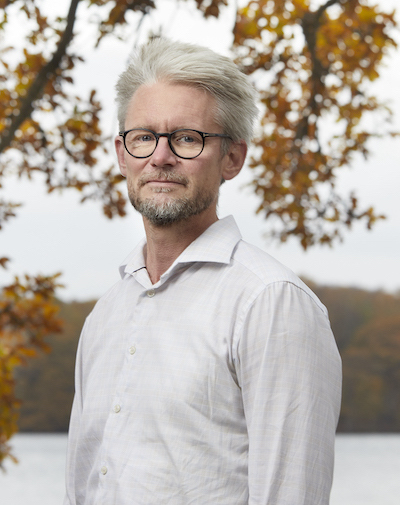}}]%
    {Tomas Bryllert} was born in Växjö, Sweden, in 1974. He received an M.Sc. degree in physics and a Ph.D. in semiconductor physics from Lund University, Lund, Sweden, in 2000 and 2005, respectively.
    
    In 2006, he joined the Microwave Electronics Laboratory, Chalmers University of Technology, Göteborg, Sweden. From 2007 to 2009, he was with the Submillimeter Wave Advanced Technology (SWAT) group, Jet Propulsion Laboratory, California Institute of Technology, Pasadena, CA, USA. He is currently with the Terahertz and Millimetre Wave Laboratory at Chalmers University of Technology, Göteborg, Sweden. He is also the co-founder and Chief Executive Officer of Wasa Millimeter Wave AB, a company that develops and fabricates millimeter wave products. Dr. Bryllert also works part-time in the new concepts team at Saab AB. His research interests include submillimeter wave electronic circuits and their applications in imaging and radar systems.
\end{IEEEbiography}

\begin{IEEEbiography}[{\includegraphics[width=1in,height=1.25in,clip,keepaspectratio]{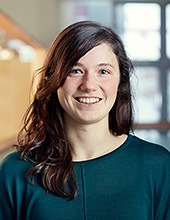}}]%
    {Marlene Bonmann} was born in Karlsruhe, Germany, in 1988. She received an M.Sc. degree in physics and astronomy and a Ph.D. in Microtechnology and Nanoscience from the Chalmers University of Technology, Gothenburg, Sweden, in 2014 and 2020, respectively.
    
    She is currently with the Terahertz and Millimetre Wave Laboratory at the Chalmers University of Technology.
\end{IEEEbiography}

\begin{IEEEbiography}[{\includegraphics[width=1in,height=1.25in,clip,keepaspectratio]{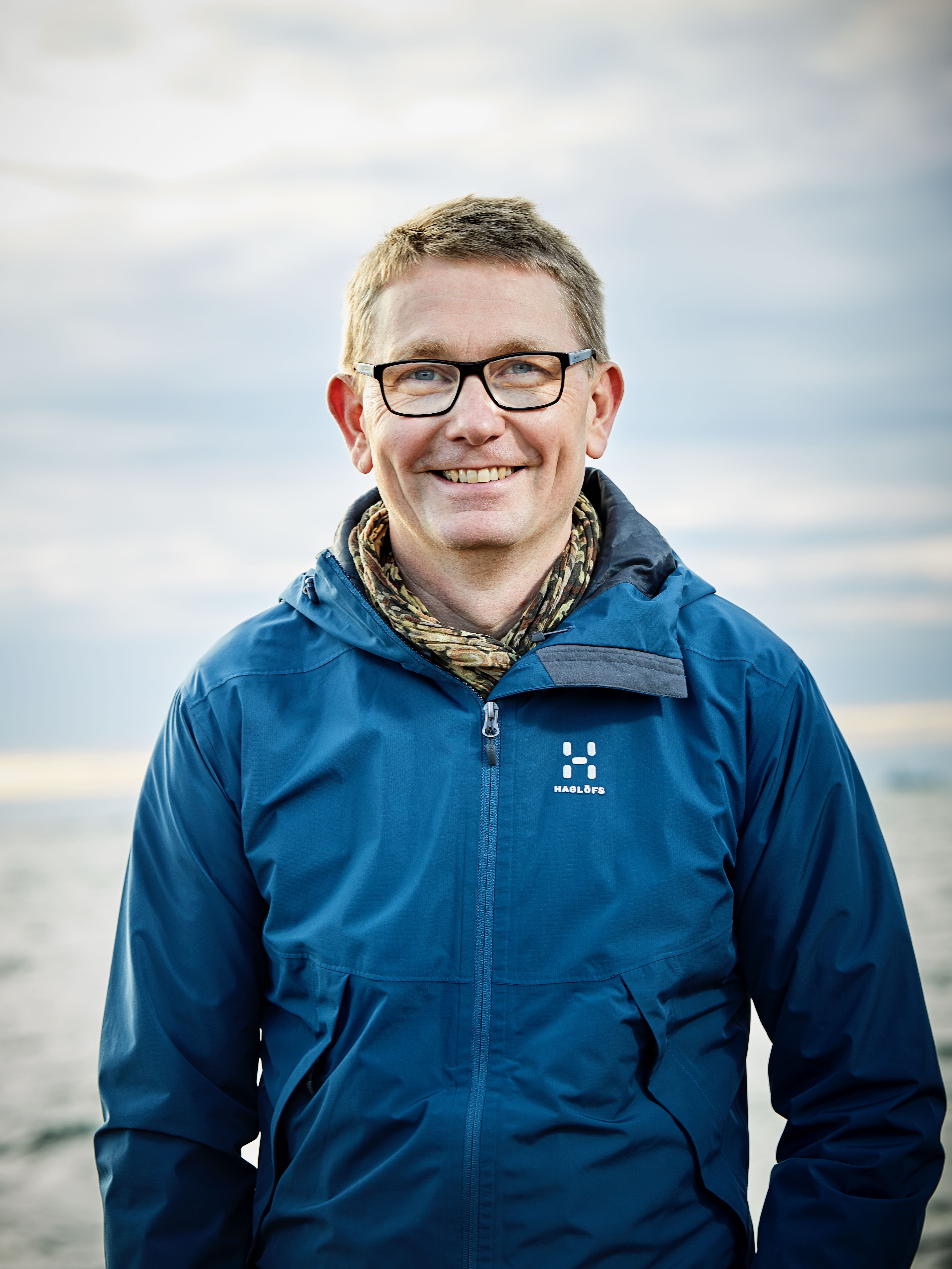}}]%
{Jan Stake} (S’95–M’00–SM’06) was born in Uddevalla, Sweden, in 1971. He received an M.Sc. degree in electrical engineering and a Ph.D. in microwave electronics from the Chalmers University of Technology, Gothenburg, Sweden, in 1994 and 1999, respectively.

In 1997, he was a Research Assistant at the University of Virginia, Charlottesville, VA, USA. From 1999 to 2001, he was a Research Fellow with the Millimetre Wave Group at the Rutherford Appleton Laboratory, Didcot, UK. He then joined Saab Combitech Systems AB, Linköping, Sweden, as a Senior RF/microwave Engineer, until 2003. From 2000 to 2006, he held different academic positions with the Chalmers University of Technology and from 2003 to 2006, he was also the Head of the Nanofabrication Laboratory, Department of Microtechnology and Nanoscience (MC2). In 2007, he was a Visiting Professor with the Sub-millimetre Wave Advanced Technology (SWAT) Group at Caltech/JPL, Pasadena, CA, USA. In 2020, he was a Visiting Professor at TU Delft. He is currently a Professor and the Head of the Terahertz and Millimetre Wave Laboratory at the Chalmers University of Technology. He is also the co-founder of Wasa Millimeter Wave AB, Gothenburg, Sweden. His research interests include graphene electronics, high-frequency semiconductor devices, terahertz electronics, submillimeter wave measurement techniques, and terahertz systems.

Prof. Stake served as the Editor-in-Chief for the IEEE Transactions on Terahertz Science and Technology between 2016 and 2018 and as Topical Editor between 2012 and 2015.
\end{IEEEbiography}
\end{document}